\documentclass[12pt,preprint]{aastex}

\usepackage{color}
\usepackage{amsmath}
\usepackage{float}
\usepackage{natbib}

\def\me{$\,{\rm M}_{\oplus}\,$}
\def\h2o{H$_2$O}
\def\sio2{SiO$_2$}
\def\gc3{g\,cm$^{-3}$}
\def\swr{Schwarzschild }
\def\ldx{Ledoux }

\begin{document}

\title{The Evolution and Internal Structure of Jupiter and Saturn with Compositional Gradients}
\author{A.~Vazan$^{1,2}$, 
R.~Helled$^{2}$, M.~Podolak$^{2}$, A.~Kovetz$^{2,3}$, \\
$^{1}$Astronomical Institute Anton Pannekoek, University of Amsterdam, The Netherlands\\
$^{2}$Department of Geosciences, Faculty of Exact Sciences, 
Tel-Aviv University, Israel\\
$^{3}$School of Physics and Astronomy, Faculty of Exact Sciences, Tel-Aviv University, Israel }

\begin{abstract}
The internal structure of gas giant planets may be more complex than the commonly assumed core-envelope structure with an adiabatic temperature profile. 
Different primordial internal structures as well as various physical processes can lead to non-homogenous compositional distributions. A non-homogenous internal structure has a significant impact on the thermal evolution and final structure of the planets. 
In this paper, we present alternative structure and evolution models for Jupiter and Saturn allowing for non-adiabatic primordial structures and the mixing of heavy elements by convection as these planets evolve. 
We present the evolution of the planets accounting for various initial composition gradients, and in the case of Saturn, include the formation of a helium-rich region as a result of helium rain.
We investigate the stability of regions with composition gradients against convection, and find that the helium shell in Saturn remains stable and does not mix with the rest of the  envelope. In other cases, convection mixes the planetary interior despite the existence of compositional gradients, leading to the enrichment of the envelope with heavy elements. 
We show that non-adiabatic structures (and cooling histories) for both Jupiter and Saturn are feasible. 
The interior temperatures in that case are much higher that for standard adiabatic models. 
We conclude that the internal structure is directly linked to the formation and evolution history of the planet.
These alternative internal structures of Jupiter and Saturn should be considered when interpreting the upcoming {\it Juno} and {\it Cassini} data. 

\end{abstract}

\section{Introduction}

The exact composition and internal structure of both Jupiter and Saturn are not perfectly known, 
but it is well-agreed that the composition of these planets is not constant with depth \citep[e.g.,][]{saum+guil04}. 
The measured physical properties of Jupiter and Saturn, in particular their masses, radii, and gravitational fields, provide constraints on their density profiles. This information, when combined with the known age of the solar-system and the measured surface temperature of the planets, constrain their thermal evolution. 
As a result, we can not only try to infer the current internal structure of the planets, but also to exclude evolution models that are inconsistent with the current luminosities of the planets. 

Typically, Jupiter is assumed to have a heavy element core surrounded by a hydrogen-helium envelope with some fraction of heavy elements \citep[e.g.,][]{hubb+militz16}. The distribution of heavy elements in the envelope is often taken to be homogeneous \citep[e.g.,][]{saum+guil04}. 
Several studies have considered a discontinuity in heavy elements between the metallic hydrogen inner envelope and the molecular outer envelope, but within each of these envelopes the heavy elements are assumed to be homogeneously distributed, and the planet is assumed to be adiabatic  \citep{guillot95,nettel12}. 
For Saturn, and recently also for Jupiter, a non-homogeneous structure was considered in the context of helium rain followed by the formation of a helium shell above the heavy-element core \citep{forthub03,hubb+militz16,pustow16}. 
However, in these models too the envelope was assumed to be homogeneous and adiabatic, simply with a lower helium mass fraction.  
   
While the standard core-envelope structure is simple 
and can reproduce the observed properties of the planets, other more complex internal structures cannot be excluded. 
Physical processes such as the miscibility of materials in hydrogen followed by core erosion \citep[e.g.,][]{guill04,wilson12a}, or planetesimal dissolution in the envelope during planetary formation \citep[e.g.,][]{iarospod07} could result in compositional inhomogeneity within the planet. 
The existence of a compositional gradient can inhibit convection \citep{ledoux47}, and lead to higher internal temperatures and a thermal profile which is non-adiabatic. 
Not only does this affect the rate at which the planet cools, but the higher internal temperatures also influence the heavy element mass fraction inferred from interior models 
\citep[e.g.,][]{chabr+barf07,lecontechab12}. 
However, even if the primordial structure is inhomogenous, efficient convection can act to homogenize the interior, and change the distribution of heavy elements as the planet evolves \citep[e.g.,][]{guill04,stevenson82a}. 
As a result, it is important to model the evolution of the planets while accounting for convective mixing. 
In \cite{vazan15}, hereafter paper~I, we introduced a planetary evolution
model that includes simultaneous heat and material transport.
In this paper, we focus on Jupiter and Saturn and investigate how different assumed primordial internal structures affect their subsequent evolution.

\section{Planetary Evolution}
To model the evolution and internal structure of Jupiter and Saturn we use a planetary evolution code (see paper~I and references therein for details). 
We use an equation of state (EOS) for a mixture of hydrogen, helium \citep{saumchab95}, and heavy-elements that are represented by \h2o or \sio2 \citep{more88}, as described in \cite{vazan13}. The opacity is derived from an harmonic mean of the radiative \citep{pollack85} and conductive \citep{potekhin99} opacities.
Convective mixing is computed self-consistently using the mixing length theory under the control of the Ledoux convection criterion (see appendix A). 

Since the primordial internal structures of Jupiter and Saturn are not well-constrained, we have considered tens of such structures, and present here a sample from the ones that result in  current-state structures that are consistent with the measured physical parameters of the planets: mass, radius, effective temperature, moment of inertia (hereafter, MOI), and the second gravitational moment (hereafter, $J_2$).
Computation methods of MOI and $J_2$ are summarized in appendix B. 
The planetary evolution is terminated at a time of 4.55 Gyr, and this is referred to as {\it the current-state internal structure}. 
It should be noted that the aim of this study is not to provide detailed current-state interior models with exact fits to the measured gravitational moments as presented in other studies \citep[e.g.,][]{nettel13,saum+guil04,miletal08,hubb+militz16}. 
Instead, our calculations  
aim to demonstrate the effect of compositional inhomogeneities on the planetary evolution, to suggest non-adiabatic internal structures for both planets, and to  demonstrate the complexity in inferring a 'final interior structure' for giant planets\footnote{In principle, it is possible to fine-tune the evolution model to achieve the exact current state radius and effective temperature. However, 
in view of the uncertainties in input physics, such as atmospheric opacities and equation of state, and the large choice of possible initial conditions, we focus at this point on investigation of plausible parameter sets rather than on converge to an exact model.}. 

\subsection{Jupiter} 

The main uncertainties in inferring Jupiter structure models are linked to the uncertainties in the hydrogen EOS, but also to the model assumptions \citep[see e.g.,][for discussion]{helled14}.  As a result, while there are limits on Jupiter's core mass and the enrichment of its envelope, their actual values cannot be said to be well-known. 
\cite{saum+guil04} have investigated the uncertainty in the inferred internal structures of Jupiter and Saturn due to the uncertainty in the hydrogen EOS and found that Jupiter's core mass can be between zero and 14\me, while the mass of heavy elements in the envelope is between 6 and 40\me. 
Internal structure models of a two-layer Jupiter using DFT-MD EOS for hydrogen and helium infer a much larger core mass of about 16\me and an envelope metallicity of 5\me \citep{miletal08}. 
A three-layer model using a different DFT EOS for hydrogen predicts a much smaller core of 0 -- 8\me for Jupiter \citep{nettel12}. 

All of these models assume that Jupiter's interior is adiabatic. 
However, Jupiter may not be fully adiabatic \citep{lecontechab12,nettel15}, and this can influence its evolution history and the inferred internal structure. In this section, we investigate the thermal and structural evolution of Jupiter for different primordial internal structures that lead to Jupiter-like planets at present.
We consider different primordial heavy-element distributions, some of them without a core in the traditional sense. 
In our models, the heavy elements are represented by \h2o. A discussion of the sensitivity of the model results to the assumed high-Z composition is given in section \ref{senseos}. 
Unless otherwise noted, the hydrogen-helium ratio is taken to be proto-solar \citep[e.g.,][]{bahcall95}.
  
\begin{figure}[ht]
\centerline{\includegraphics[angle=0, width=15cm]{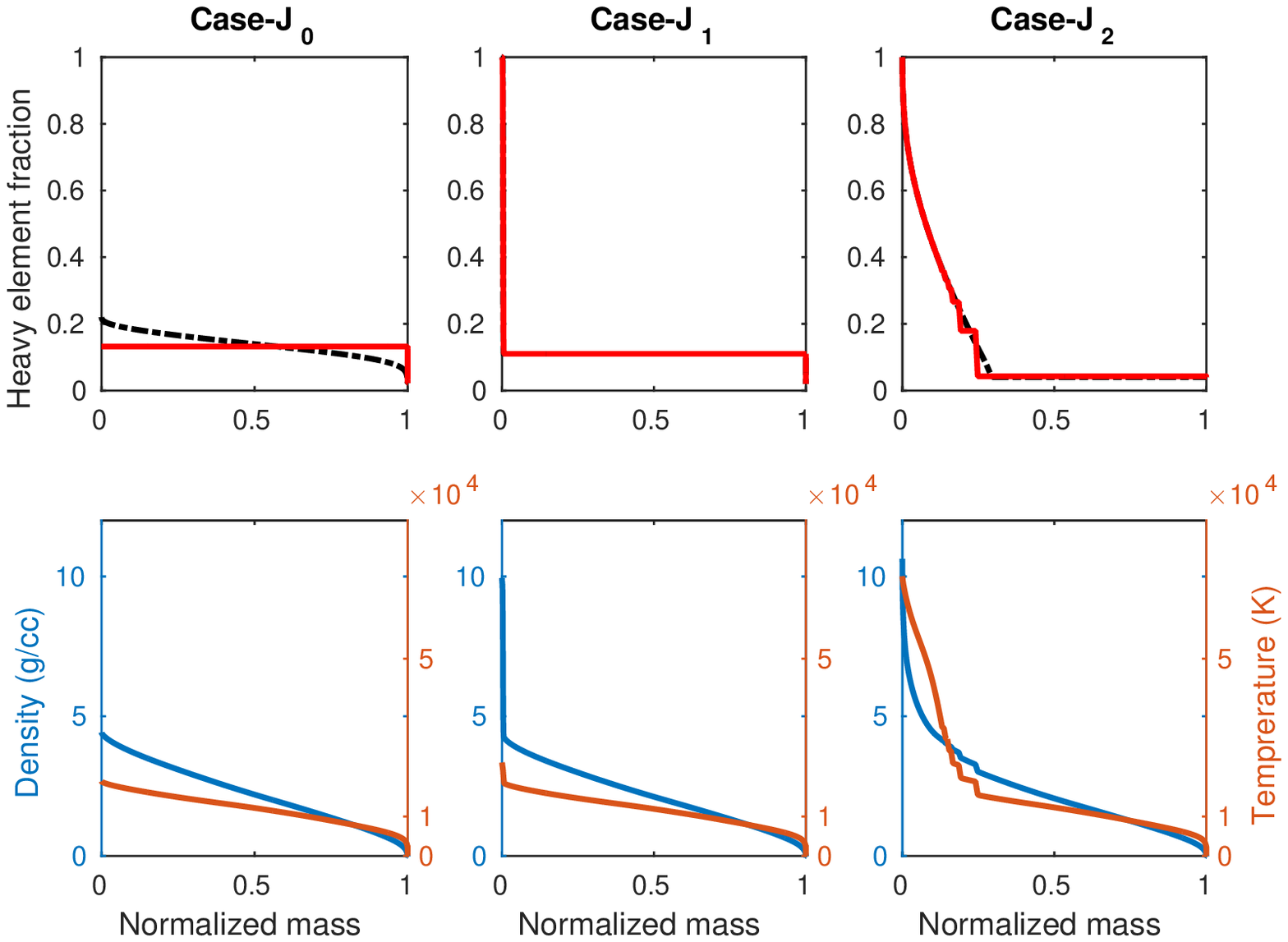}}
\vspace{-10pt}
\caption{{\bf Top:} The primordial (dashed-black) and current-state (solid-red) distribution of heavy elements in Jupiter for Case-J$_0$ (left), Case-J$_1$ (middle) and Case-J$_2$ (right). 
{\bf Bottom:} The density (blue) and temperature (red) profiles for the current-state internal structure for Case-J$_0$ (left), Case-J$_1$ (middle) and Case-J$_2$ (right).}
\label{jupZ}
\end{figure}

In the first case, Case-J$_{0}$, which is shown in Figure~\ref{jupZ}, the heavy-element distribution is moderate, i.e., the mass fraction $Z$ of heavy elements  does not exceed about 0.2, similar to the distribution suggested by \cite{lecontechab12}.
The compositional gradient of Case-J$_{0}$ is insufficient to inhibit convection, and convective mixing leads to a homogeneous composition across the envelope within a few $10^7$ years. 
The initial temperature profile determines the rate of mixing, so that lower initial temperatures result in convection and mixing on longer timescales. 
Since the primordial internal structure of Case-J$_{0}$ is not maintained on a long timescale, and the planet becomes fully convective (aside from an outermost radiative region), we suggest that layered convection is unlikely to occur for this configuration. 
The initial configuration of Case-J$_{0}$ could fit Jupiter's measured properties if the compositional gradient remains stable during the evolution. However, our simulations suggest that because of the mixing, the current-state structure (figure~\ref{jupZ}, red curve in upper left panel) cannot reproduce Jupiter's measured $J_2$ moment (see Table~\ref{tab1}). Other core-envelope structures that fit the observations do exist, as shown in the example of Case-J$_{1}$. In this case, Jupiter has a small core with a mass of 2\me, and an envelope with Z=0.11 which is adiabatic.   

In Case-J$_{2}$ in figure~\ref{jupZ}, the initial heavy-element gradient is much steeper than in Case-J$_{0}$, decreasing gradually from $Z=1$ in the center. 
Here, the innermost regions (inner $\sim$15\% of the mass) are found to be stable against convection, and therefore, act as a bottleneck in terms of heat transport, while the outer envelope is convective throughout the entire evolution. 
The increasing temperature gradient between the innermost non-adiabatic and non-convective region and the outer convective region allows the convective region to slowly penetrate inward during the  evolution, and this leads to a small enrichment in heavy elements in the outer envelope as time progresses. 
The steep composition gradient in the innermost region inhibits convection, and layered convection can occur (see section \ref{semi} for details). This could modify the steep temperature gradient. 
The top panels in Figure~\ref{jupZ} show the heavy element distributions for the initial (dashed) and current-state (solid) configurations, while the lower panels show the current-state temperature (red) and density (blue) profiles. It is interesting to note that although both Jupiter models (Case-J$_{1}$ and Case-J$_{2}$) have the same mass and similar composition, the internal structures and temperatures are quite different.  
The physical properties of the current-state internal structures for the two cases are listed in Table~\ref{tab1}. 

\begin{figure}[ht]
\centerline{\includegraphics[angle=0, width=22cm]{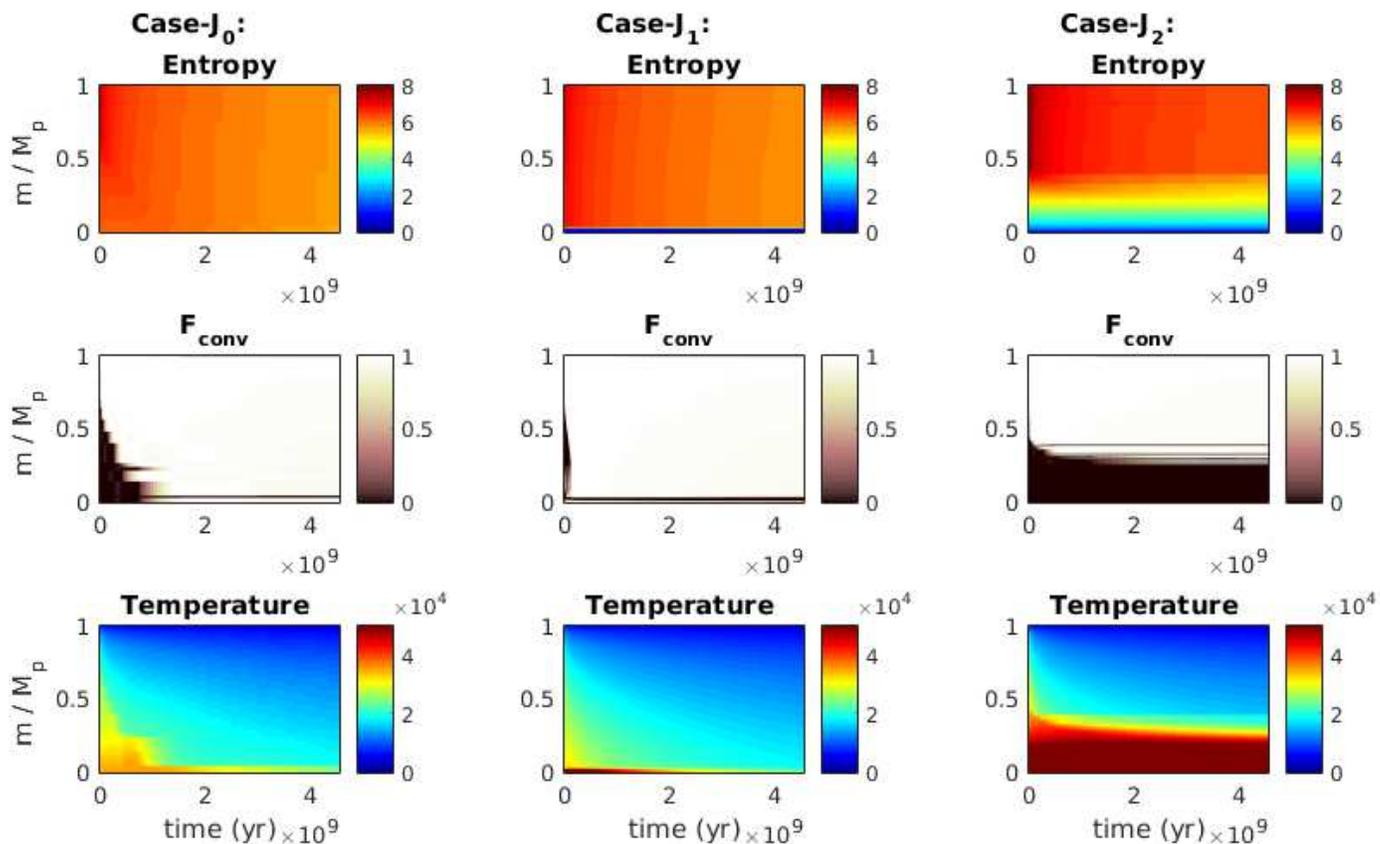}}
\vspace{-10pt}
\caption{Thermal evolution for Case-J$_0$ (left), Case-J$_1$ (middle) and Case-J$_2$ (right). Shown are the log entropy (top), convective efficiency (middle), and temperature (bottom), as a function of time (x -axis) and normalized planetary mass (y-axis). The entropy is given by specific entropy per Baryon units, the convective efficiency is the fraction of energy that is transported by convection, and the temperature is in degrees Kelvin.} 
\label{sjup}
\end{figure}

The evolution of the models is shown in Figure~\ref{sjup}. The colors represent the specific entropy (upper panel), the fraction of energy transferred by convection (middle panel), and the temperature (lower panel) profiles, as a function of normalized planetary mass  (y-axis) and time (x-axis). 
Since in Case-J$_0$ the planet becomes homogenous after several million years, the energy is transferred via convection (white color in the middle panel) and the entropy is nearly constant throughout the interior as expected from an adiabatic structure, similar to Case-J$_1$. 
In Case-J$_0$ and Case-J$_1$, assuming an adiabatic structure is appropriate. 
In Case-J$_2$ the innermost region which is highly enriched in heavy element has a lower entropy and is stable against (large-scale) convection during the entire evolution (dark regions, middle panel). As a result, the temperatures in the inner regions remain high while the outer ones can cool efficiently.

\subsection{Saturn}

For Saturn too, uncertainties in the EOS result in corresponding uncertainties in the inferred composition and core mass. Unlike the case for Jupiter, most modelers agree that Saturn must have a sizeable heavy element core in order to fit the measured gravitational field. 
\cite{saum+guil04} found a 10 -- 25\me core and 1 -- 10\me envelope enrichment, while \cite{nettel13} infer a core mass between 0 and 20\me, and an atmospheric enrichment range of 5 -- 15\me. 
A recent study of the uncertainty in Saturn's internal structure due to the uncertainties in planetary shape and rotation rate suggested a core mass of 5 -- 20\me, and 0 -- 7\me of heavy elements in Saturn's envelope \citep{hellguill13}. 

\begin{figure}[ht]
\centerline{\includegraphics[angle=0, width=15cm]{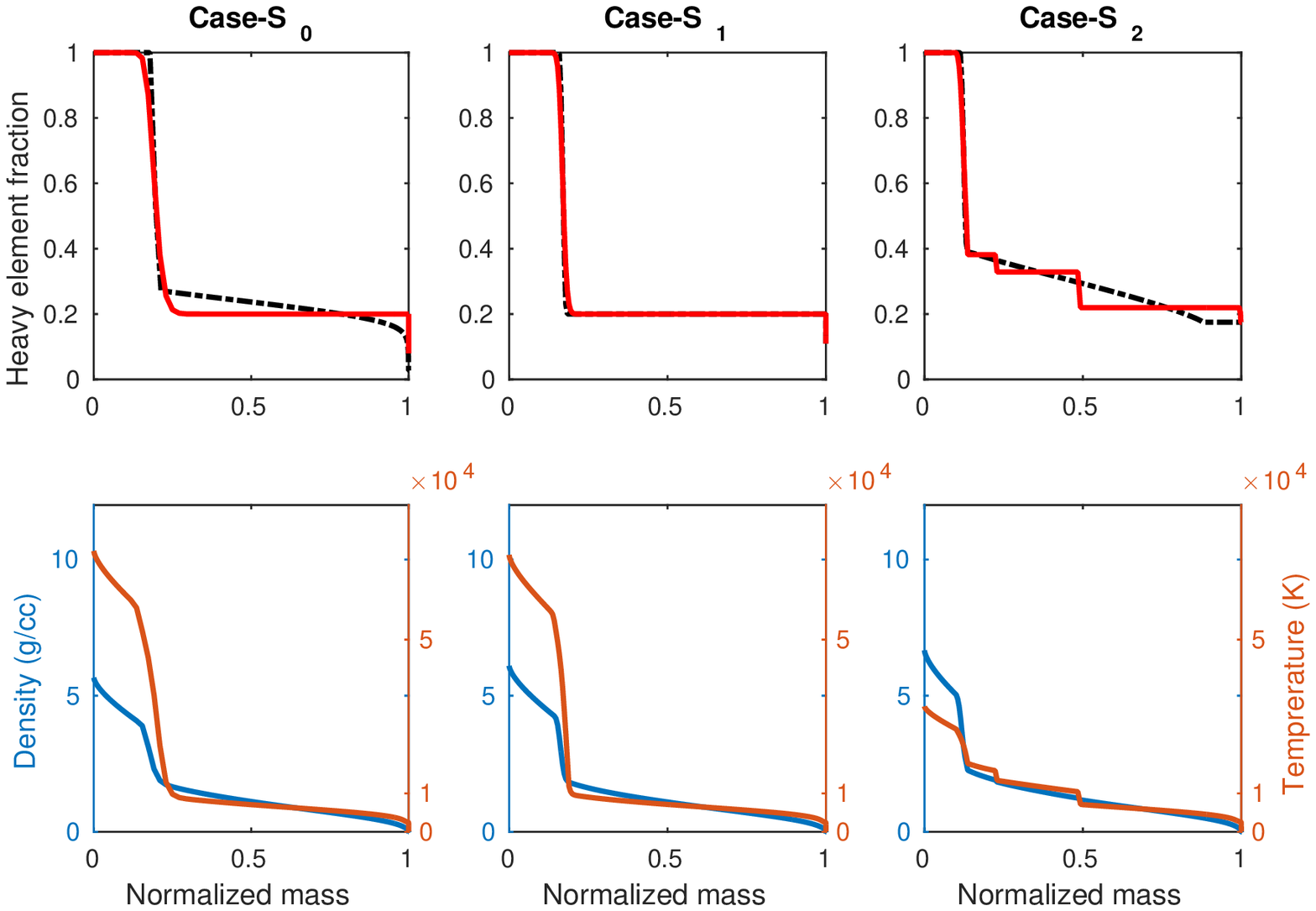}}
\vspace{-10pt}
\caption{{\bf Top:} The primordial (dashed-black) and current-state (solid-red) distribution of heavy elements in Saturn for Case-S$_0$ (left), Case-S$_1$ (middle) and Case-S$_2$ (right). 
{\bf Bottom:} The density (blue) and temperature (red) profiles for the current-state internal structure for Case-S$_0$ (left), Case-S$_1$ (middle) and Case-S$_2$ (right).}
\label{strnZ}
\end{figure}

Saturn's current luminosity is larger than predicted from homogeneous and adiabatic evolution models \citep{stevsal77a,pollack77,fort+nettel10}.
A natural explanation is a non-adiabatic evolution caused by helium rain \citep{stevsal77a}. 
However, Saturn's high luminosity could also be a result of heavy element gradients \citep{lecontechab13}. 
In order to further investigate this question, we first present evolution models of Saturn in which the compositional gradients are in the heavy elements alone (i.e., no helium rain). The results are presented in the upper panel of Fig.~\ref{strnZ}. 
Case-S$_{0}$ has a total heavy-element mass of 34\me, where 19\me are in the core and the rest of the heavy-element mass is gradually distributed starting from Z=0.3 on top of the core to Z=0.04 in the outermost regions \citep[similar to the case of][]{lecontechab12}.
Case-S$_1$ represents a traditional core-envelope model that fits Saturn's observed properties.  
In Case-S$_{2}$, the composition gradient is steeper, with Z=0.4 just above the 12\me core and decreasing outward to Z=0.18 in the outermost envelope. 
For Saturn too, we have modeled many cases with different compositional gradients and initial conditions, but we present here a sample from the ones that are consistent with Saturn's measured properties at present. 
 
\begin{figure}[ht]
\centerline{\includegraphics[angle=0, width=20cm]{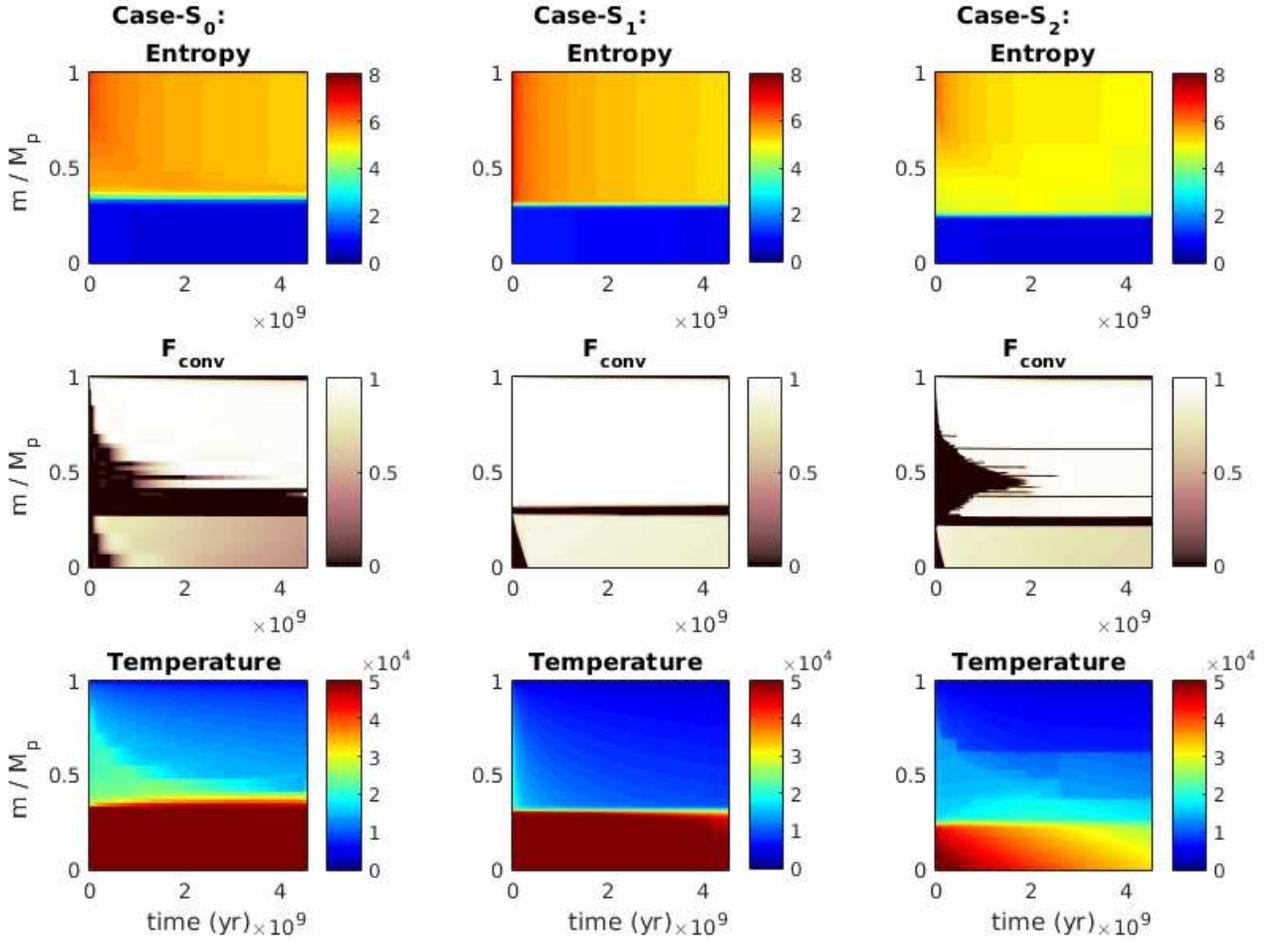}}
\vspace{-10pt}
\caption{Same as Figure 2 for the Saturn cases: Case-S$_0$ (left), Case-S$_1$ (middle) and Case-S$_2$ (right).}
\label{sstrn}
\end{figure}

The initial (dashed) and current-state (solid) heavy-element distributions for the cases are shown in Fig.~\ref{strnZ}. Similarly to Jupiter, a moderate compositional gradient (Case-S$_{0}$) cannot be maintained after $\sim$10$^7$ years, and the envelope becomes homogeneous, leading to a convective and adiabatic envelope. 
After 4.55 Gyrs of evolution, this model cannot reproduce Saturn's measured physical properties. 
Interestingly, although Case-S$_1$ is a simple core-envelope model, it can reproduce the measured effective temperature of Saturn. 
In this case the evolution begins with a hot internal structure (high initial energy content), and the core-envelope boundary acts as a bottleneck for the heat transport.
As a result, the core's temperatures are much higher than in the standard adiabatic model, and the (almost) adiabatic envelope can still match Saturn's observed properties, including its effective temperature. 
In contrast to sharp core-envelope boundary, a steeper heavy element gradient (Case-S$_2$) inhibits large-scale convection, and the heat is retained in the inner region. 
As a result, in this case lower (than Case-S$_1$) internal temperatures are needed in order to reproduce Saturn's observed properties. 
Nevertheless, in all cases, in order to reproduce the estimated value of Saturn's MOI, a relatively massive core is required. This result is in good agreement with previous Saturn structure models \cite[e.g.,][]{nettel12,hellguill13,saum+guil04}.
The evolution of Saturn for the cases is presented in Fig.~\ref{sstrn}.
In Case-S$_0$ (left), convection penetrates into the innermost regions during the first few millions of years, leading to a fully convective envelope, while the homogeneous envelope of Case-S$_1$ (middle) is fully convective from the very beginning. 
The core-envelope boundary, however, remains radiative (dark color). 
In Case-S$_2$, the envelope develops several convective regions (separated by thin radiative layers) during the evolution (dark color), which leads to a moderate temperature profile in the current-state model. 
In all of these cases the temperatures at the innermost region must be high (than usually assumed in interior models) in order to reproduce the observed high luminosity of Saturn. Therefore, we suggest that in principle, Saturn can have an extended outer region which is adiabatic and convective as long as its primordial internal structure is hot (see Table~\ref{tab1} and Figure~\ref{sstrn}).

If helium separates from hydrogen, one must account for this effect when modeling Saturn's evolution. The occurrence of helium rain in Saturn is consistent with the depletion of helium in its atmosphere \citep[e.g.,][]{congaut00,guillgaut14}, and may explain the low MOI of Saturn as well as its slow cooling \citep[e.g.,][]{forthub03}. 
In addition, helium rain could lead to the formation of a helium shell above the heavy-element core \citep[e.g.,][]{stevsal77a,forthub03} 
that can not only lead to a condensed-Saturn without the need for a massive heavy-element core \citep{forthub03} 
but also to the creation of another compositional boundary within the planet. 
We next investigate Saturn models in which helium rain is included by adding helium-rich regions above the heavy-element core. 
According to calculations of the helium-hydrogen phase diagram, under certain pressure-temperature conditions the miscibility of helium in hydrogen is reduced \citep[e.g.,][]{pfaffenz95,morales09}. As a result, part of the helium in Saturn (and Jupiter) separates from hydrogen as droplets, and settles into the inner regions, even in convective regions \citep{stevsal77b}. The settling timescale is predicted to be short, and therefore helium-rain is typically assumed to happen instantaneously in planetary evolution models \citep{forthub03}. 

In this work, we relocate the helium in a shell above the core when the pressure-temperature conditions in the planet enters the demixing region, according to the phase diagram of \cite{morales09}. The helium droplets are expected to redissolve in hydrogen when they leave the immiscibility region, and therefore, the helium shell in our model is not of a pure helium but is mixed with some hydrogen. The thickness of the shell and its location are similar to the ones presented by  \cite{stevsal77a,forthub03}.  
The relocation of helium in deeper layers results in a drop in entropy in the planet's helium-rich region, as in \cite{forthub03}. A new entropy profile is calculated for the new composition distribution by using our mixture EOS, as is described in appendix A1 in \cite{vazan13}.
For this entropy profile we calculate energy, temperature and density profiles. These are derived using the planetary evolution code. After a new structure is defined, we continue the evolution from this point.
As a result of helium settling, the energy budget of the planet is changed, and gravitational energy is released as thermal energy in the helium-rich region.

\begin{figure}[ht]
\centerline{\includegraphics[angle=0, width=15cm]{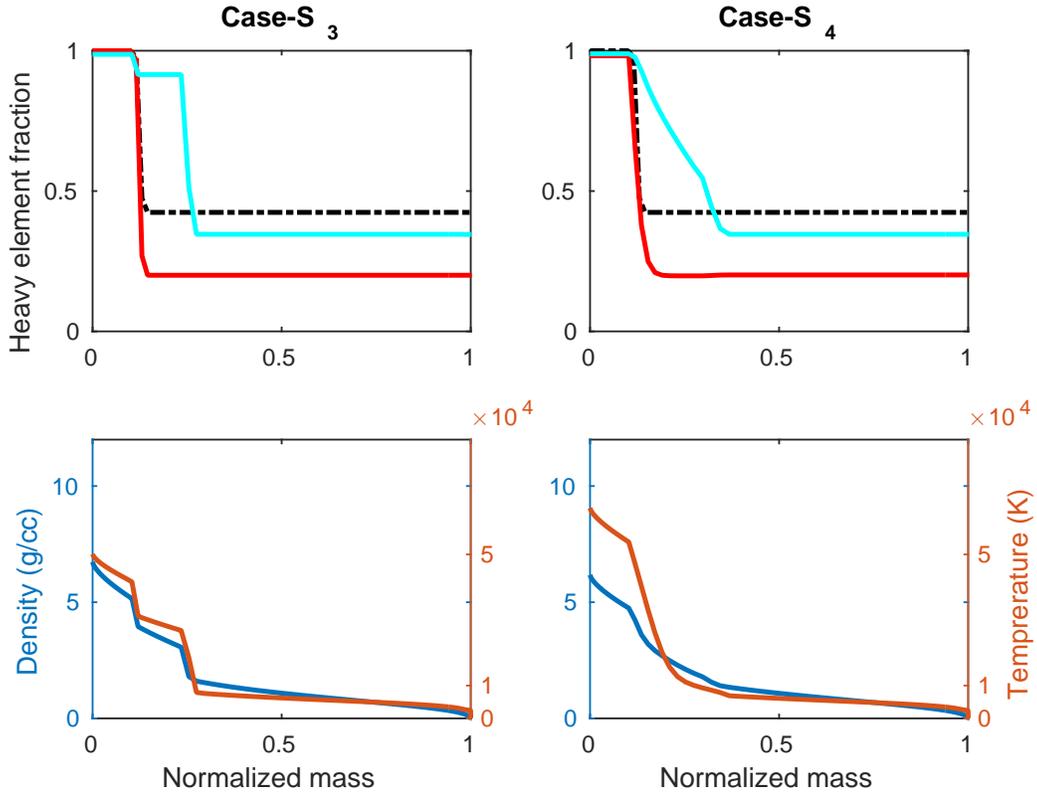}}
\vspace{-10pt}
\caption{{\bf Top:} The primordial distribution of heavy elements and helium (dashed-black) and current-state distribution of heavy element (solid-red)  and of helium (solid-cyan) in Saturn for Case-S$_3$ (left) and Case-S$_4$ (right). 
{\bf Bottom:} The density (blue) and temperature (red) profiles for the current-state internal structure for Case-S$_3$ (left) and Case-S$_4$ (right). }
\label{strn2Z}
\end{figure}

In Case-S$_{3}$, the primordial model has a 12\me heavy-element core surrounded by an envelope with $Z=0.20$. 
The helium shell is added to the model after 2.9 Gyr of evolution. 
In Case-S$_{4}$, we use the same primordial model as in Case-S$_3$ with the difference that the helium-shell is assumed to have a gradual distribution of helium. In both cases the helium in the shell is mixed with some hydrogen, as presented in the upper panels of Fig.~\ref{strn2Z} (cyan curve), and the total helium abundance in the planet is held constant (before and after the helium settling). 
The results for the two cases with helium settling are shown in Fig.~\ref{strn2Z}. 
In these cases the helium shell remains stable against convection throughout the reminder of the planetary evolution. Since helium separation occurs relatively late, the temperature gradient in the envelope is not steep enough to initiate convective mixing and remix of the helium shell.  
The helium shell forms an additional composition boundary 
which retains the heat. 

\begin{figure}[ht]
\centerline{\includegraphics[angle=0, width=17cm]{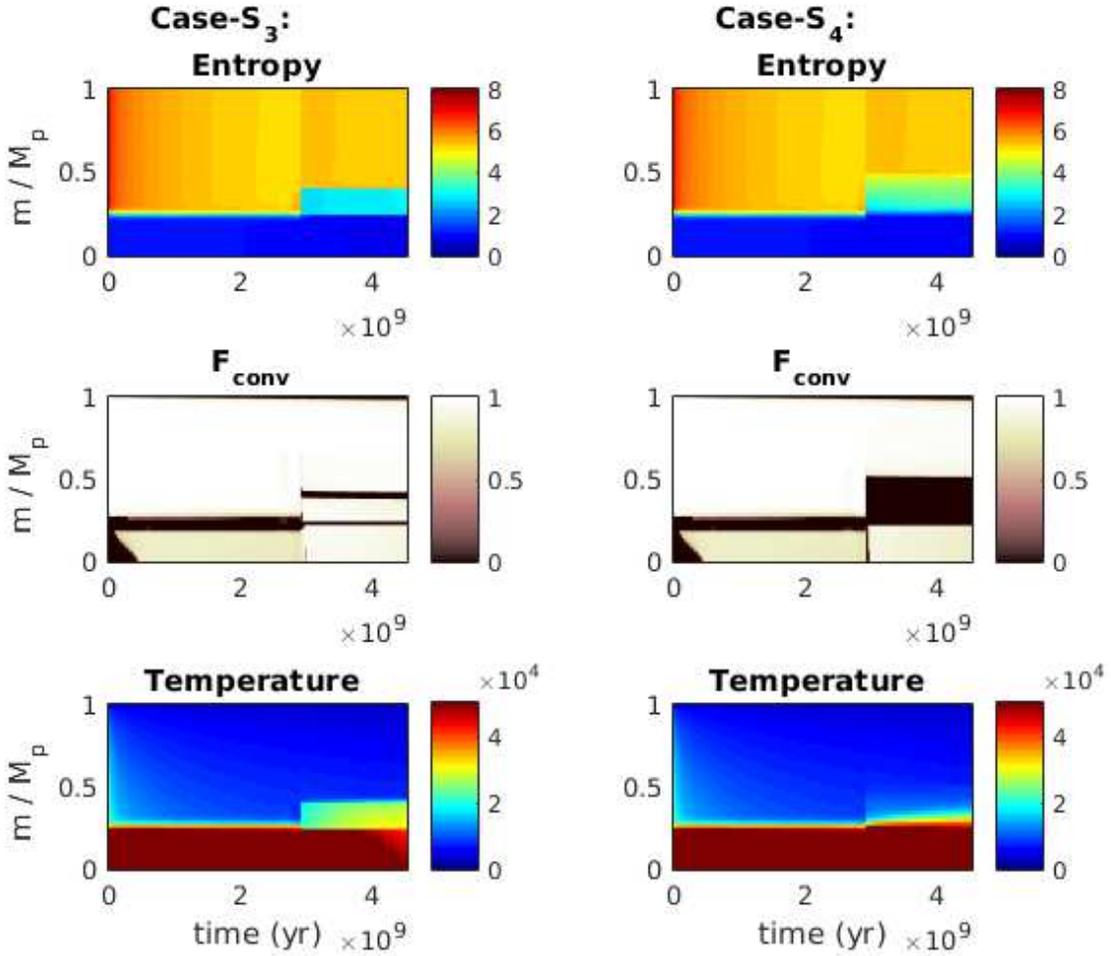}}
\vspace{-10pt}
\caption{Thermal evolution for Case-S$_3$ and Case-S$_4$. Also here, shown are the entropy (top), convection efficiency (middle), and temperature (bottom) as a function of time (x -axis) and normalized planetary mass (y-axis). The discontinuity in the model occurs due to the inclusion of a helium shell (see text for details). Units as in figure~\ref{sjup}.}
\label{sstrn2}
\end{figure}

The entropy, temperature, and density evolution for the two cases are shown in Fig.~\ref{sstrn2}. 
As shown in the lower panels of Fig.~\ref{sstrn2}, the increase in temperature affects the region which is enriched with helium, but has a negligible effect on the outer temperatures. It should be noted that gravitational energy is expected to be released in this process and converted into thermal energy in a gradual manner, while in this work the change in energy is instantaneous. 
The sharp composition boundary that is formed prevents the planet from releasing the heat associated with this process efficiently. 
This can lead to differences in the inferred internal temperatures. 
However, the effect of this heat pulse on the subsequent evolution is small. The decrease in gravitational energy results in a thermal energy increase by 6\% and 8\% for Case-S$_{4}$ and Case-S$_{3}$, respectively.
This increases the shell's temperature by a factor of about three right after the helium settles, but has a negligible effect on the total luminosity of the planet.  
Since the rate of heat transport from the helium shell essentially determines the contribution of the process of helium rain to the evolution of the planet, the current-state effective temperature is also affected by the timescale of the heat release. 
Slow heat transport, due to compositional gradients in this case, has a smaller (but ongoing) effect on Saturn's effective temperature. 
We suggest that while the helium shell is important for explaining the detected low helium abundance in Saturn's atmosphere, and can reproduce Saturn's MOI and $J_2$ moment with a small core, its effect on the thermal evolution might not be sufficient to explain the high luminosity of Saturn.

\begin{figure}[ht]
\centerline{\includegraphics[angle=0, width=15cm]{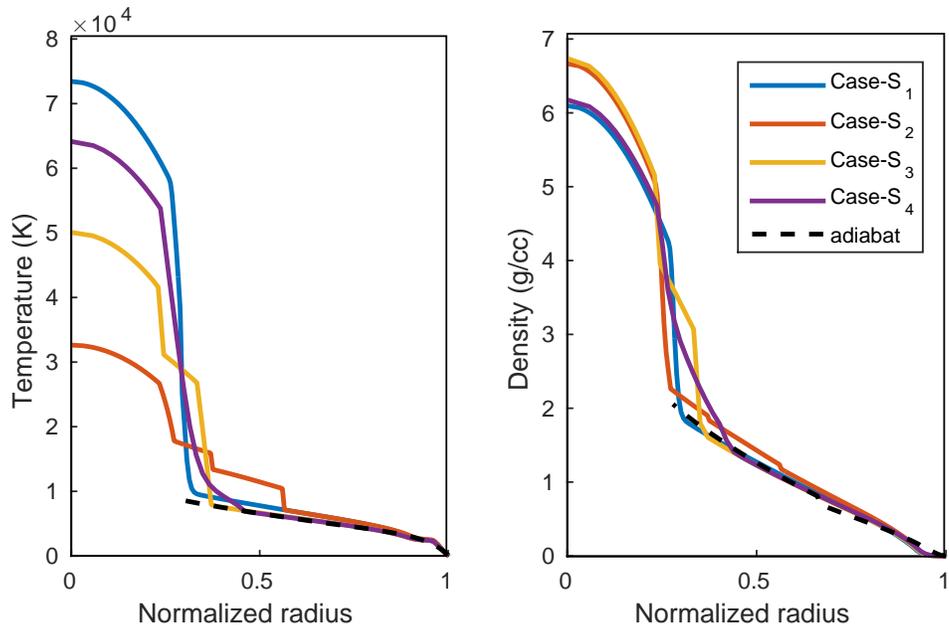}}
\vspace{-10pt}
\caption{The temperature (left) and density (right) profiles for the current-state internal structure for the four Saturn models. For comparison, also shown is an adiabatic envelope model of Saturn \citep{hellguill13}.}
\label{strnT}
\end{figure}

While the last four Saturn models we present can reproduce Saturn's measured physical properties relatively well, the internal density and temperature profiles can differ significantly. The temperature (left panel) and density (right panel) profiles for the current-state internal structures are shown in Fig.~\ref{strnT}.
For comparison, we also present an adiabatic envelope structure (dashed-black) taken from \cite{hellguill13}. In our models, the internal temperatures are typically higher than the standard adiabatic case. 
Case-S$_1$ in which the envelope is almost fully-convective actually has the highest internal temperatures. This is linked to the fact that the initial temperature profile (the energy content of the initial model) must be high in order to provide the proper observed luminosity at the present time. 
In Case-S$_2$ the small composition "steps" inhibit large scale convection and the subsequent heat transport, and therefore core temperature does not need to be as high as in Case-S$_1$.
The helium shell above the core for Case-S$_3$ leads to discontinuities in the temperature profile due to the formation of composition boundaries (core-helium, helium-envelope) where convection is inhibited. The gradual distribution of helium shell in Case-S$_4$, results in a gradual temperature profile in the helium-rich region. 
The physical properties of the current-state structure for the Saturn models are listed in Table~\ref{tab1}. 
\par

\subsection{The Sensitivity of the Results to Model Assumptions}\label{accur}

\subsubsection{EOS}\label{senseos} 
\begin{figure}[ht]
\centerline{\includegraphics[angle=0, width=15cm]{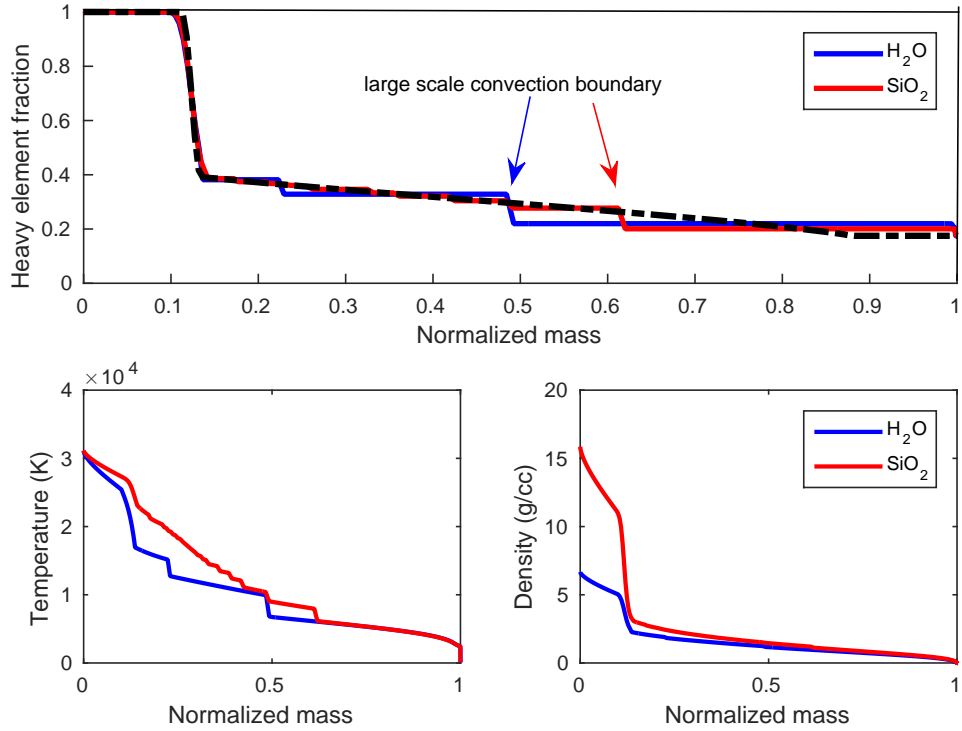}}
\vspace{-10pt}
\caption{{\bf Top:} The primordial (dashed-black) and current-state (solid) distributions of heavy elements in Saturn for Case-S$_2$ when using H$_2$O (blue) and SiO$_2$ (red). The arrows mark the outer region in the current-state internal structure that is fully-convective. {\bf Bottom:} Temperature (left) and density (right) for the current-state internal structure with the heavy elements being represented by H$_2$O (blue) and SiO$_2$ (red).}
\label{strnsh}
\end{figure}

The calculated evolution and structure models correspond to a specific heavy element composition and its EOS. 
In the models presented above, the heavy elements are represented by \h2o, and therefore, the derived core density is relatively low (up to 11 g cm$^{-3}$ for Jupiter and 7 g cm$^{-3}$ for Saturn).
In order to examine the sensitivity of the results to the assumed heavy element composition we also ran models in which the high Z material is rocky (\sio2).  
When the core material is taken to be SiO$_2$, using an EOS based on \cite{qeos88} (see paper~I for details), the central densities can be as high as 22 g cm$^{-3}$ for Jupiter and 15-20 g cm$^{-3}$ for Saturn, depending on the model.
The radii of the planets, are smaller by up to 7\% for the Jupiter and Saturn models we have considered. The moment of inertia (MOI) varies slightly as well (several percents).
A comparison between \h2o and \sio2 for Case-S$_2$ is shown in Figure~\ref{strnsh}. 
Due to its lower molecular weight, water mixes more easily, and as a result, half of the planetary mass (outer region) in the current-state model is fully mixed in comparison with only 40\% of the mass for the case of \sio2 (see paper I for more details). 
In addition to differences in density, the temperature profile is also affected with the temperatures in the inner envelope being higher by up to 50\% in the case of SiO$_2$. 
In the core, however, the difference is found to be small.
It should be noted that when modeling Case-S$_2$ with an EOS of \sio2, the observed parameters of Saturn cannot be reproduced. This model is presented here in order to demonstrate the effect of the  heavy element EOS on the calculation. 
For other compositions of the heavy elements, different mass ratios and/or different compositional gradients would be required in order to reproduce the measured properties of Jupiter and Saturn. 

\subsubsection{Opacity}
The planetary evolution is computed using a specific atmospheric boundary condition, which is taken to be the planetary photosphere. 
I.e., the outer boundary conditions is taken as,  
\begin{equation}
\kappa p = \tau_s g,
\end{equation}
where $\kappa$ is the opacity, $p$ is the pressure, $g=GM/R^2$ is the gravitational acceleration, and $\tau_s$ is the optical depth of the photosphere. 
The temperature at the 1-bar pressure level is affected by the choice of the opacity source and by the temperature profile. 
The default radiative opacity calculation in our models \citep{pollack85}  does not  reproduce the temperature measured for Jupiter's 1-bar level.  This is not surprising, given that this opacity table includes small interstellar grains which are 
not expected to exist in the upper atmosphere of the giant planets due to grain settling and cloud formation. As a result, in order to  reproduce the correct temperatures at 1-bar we must assume a grain-free atmosphere. 
For the grain-free atmosphere we simply use the Rosseland mean of the gas opacity of \cite{sharp07}.  
A proper determination of the opacity in the giant planet is important also for identifying the convective regions within the planet (e.g., Guillot et al., 1995). 
A self-consistent calculation of the local opacity as a function of the local metallicity during the planetary evolution is desirable, and 
we hope to address that in future work.

\subsubsection{The Planetary Albedo and Solar Irradiation}
The albedos of Jupiter and Saturn were taken to be 0.343 and 0.344, respectively \citep{forthub03}. While these values correspond to the current-state of the planets, it is unclear how the albedo changes as the planets evolve. 
For simplicity, and given that there are no constraints on the planetary albedos at early times, we set the albedos of both planets to be constant in time. However, it is not unlikely that the planetary albedo changes with time due to chemical interactions and physical processes taking place in the atmosphere. 
The value of the albedo essentially determines the thermal evolution and the effective temperature of the planet. 
For Saturn, if we use a much smaller value for the albedo ($\sim$ 0.1) during most of the evolution, the correct effective temperature can be reproduce without including non-convective regions within the planet. At the moment, it is unclear whether such a low-value can be justified, and it would be interesting in future research to investigate physical/chemical mechanisms that can affect the planetary albedo at early stages. 
Another important parameter that is used in the model is the irradiation from the Sun. The equilibrium temperatures of Jupiter and Saturn are 110 K and 81 K, respectively \citep{forthub03}.
As expected from stellar evolution models \citep[e.g.,][]{Mowlavi12},  only small changes in the stellar luminosity are expected during the long-term evolution (less than 1.5\% between 1-4.55 Gyr). Therefore, the equilibrium temperatures with the Sun can be taken as constant during the long-term evolution.  

\subsection{Layered-Convection} \label{semi}
In our evolution model, the presence of convection in regions with compositional gradients is determined by the ratio between the destabilizing temperature gradient and the stabilizing composition gradient; if the latter is dominant - the heat is assumed to be transferred by radiation and/or conduction.
Regions with high fraction of heavy elements that are found to be radiative in our model could in principle develop layered-convection \citep[e.g.,][]{rosenblum11, wood13}. 
Layered-convection is expected to occur in regions that are found to be stable against convection when considering the \ldx criterion, but unstable for the \swr criterion (see appendix A). 
In these regions the heat transport rate in our calculations is lower than in the case of layered-convection, since we treat these regions as being radiative/conductive. 
While layered-convection could be an important phenomenon in giant planet evolution, it seems to be limited to specific cases. We suggest that layered-convection in Jupiter and Saturn \citep{lecontechab12} is possible only when having steep initial gradients of the heavy elements.  

For the cases with regions of steep compositional gradients where layered-convection can occur, the heat transport and the onset of convection, can differ from the ones presented here. Thus, our models provide a lower bound for the thermal cooling and the efficiency of heavy-element mixing (see paper~I).
Modeling the planetary evolution accounting for layered-convection is desirable and we hope to address this topic in a future work.
Including layered-convection in planetary evolution models, however is non-trivial since it requires knowledge about the Nusselt and Rayleigh numbers, which depend on physical properties such as the thermal and molecular diffusivities, which are not well known \citep[see e.g.,][]{mirouh12}. In addition, the heat transport and inferred composition in that case also depend on 
the assumed number of convective-diffusive layers \citep{lecontechab12}. As a result, it is very difficult to make a clear prediction for how the presence of layered-convection would affect the evolution of Jupiter and Saturn.

\section{Conclusions and Discussion}
In this work, we suggest non-adiabatic and non-homogenous evolution and structure models of Jupiter and Saturn, and investigate how the choice of the primordial internal structure and bulk composition affect the evolution of the planets.
We present various primordial heavy-element distributions for which the planetary evolution reproduces fairly well the measured physical parameters at present. 
As our models provide good, albeit not 
exact matches to the observed parameters of Jupiter and Saturn, we conclude that both planets can, in principle, have non-convective regions, and as a result, much hotter interiors. 

We show that a moderate primordial heavy-element gradient (e.g., Case-J$_0$ for Jupiter and Case-S$_0$ for Saturn) becomes homogenous via convective mixing after several million years. This mixing results in an enrichment of the planetary envelope with heavy elements. On the other hand, if the primordial composition gradient is steep (e.g., Case-J$_2$ and Case-S$_2$ for Jupiter and Saturn, respectively) convection in the deep interior is inhibited. This affects the thermal evolution, and leads to hotter interiors in comparison to the standard adiabatic case.
In these cases the innermost regions retain heat while the outer envelope cools by convection, and the convective region expands inward as time progresses leading to an enrichment of the envelope with heavy elements. As a result, convective mixing should be considered as a possible mechanism for increasing the atmospheric metallicities in giant planet atmospheres \citep[e.g.,][]{guillgaut14}. For Saturn, we consider additional cases with a helium shell above the heavy-element core when helium separation is expected to occur. 
The helium shell, whether it is distributed homogeneously (Case-S$_3$) or gradually (Case-S$_4$), is found to remain stable during the evolution and does not mix with the helium-poor regions. If the formation of a helium-rich region is fast we argue that helium rain alone is insufficient to reproduce Saturn's  high luminosity.

Our findings suggest that the initial configuration has an important role in determining the long-term planetary evolution (see also paper~I). If the  primordial internal structure is hot, the planet is expected to retain heat even after 10$^9$ years of evolution. 
The importance of the initial conditions on the current-state is best demonstrated in Case-S$_1$ for Saturn. In that case the core-envelope internal structure is consistent with the measurement of Saturn's effective temperature, due to its hot initial configuration. 
{\it It is therefore clear that constraining the primordial internal structure of giant planets shortly after their formation is crucial for the investigation of their evolution histories and internal structure. }

The very different, but possible, internal structures and evolutionary tracks of Jupiter and Saturn should be considered when 
interpreting the upcoming {\it Juno} and {\it Cassini} data. 
The existence of non-adiabatic structures which results in higher internal temperatures, can lead to a very different predicted heavy-element and core masses. In addition, different primordial internal structures result in different mixing patterns, and therefore  different enrichment in heavy elements in the atmosphere.

Finally, our results also relevant for the characterization of giant exoplanets. While typically the radius of a given planetary mass decreases with an increasing heavy-element mass, the possibility of non-convective internal structures, directly affect the cooling of the planet, and therefore, its radius as a function of time. This demonstrates the importance of determining the age of the planet accurately. This should then be combined with knowledge of the primordial structure and/or initial conditions based on planet formation models. 
Even for the giant planets in the solar system, whose internal structures are much more constrained by various measurements from space missions, it is not yet straightforward to constrain their bulk compositions and to determine their evolutionary paths.  

\section*{Acknowledgments}
We thank an anonymous referee for his/her valuable comments.
We thank Carsten Dominik and Nadine Nettelmann for helpful remarks.
A.~V.~acknowledge support from the Israeli Ministry of Science via Ilan Ramon fellowship.
R.~H.~acknowledges support from the Israel Space Agency under grant 3-11485, and from the United States - Israel Binational Science Foundation (BSF) grant 2014112. 

\section*{APPENDIX}
\appendix
\section{Convection with heavy-element gradients}
The onset of convection when accounting for heavy-element gradients is determined by the Ledoux's criterion \citep{ledoux47},
\begin{equation}
\nabla_R - \nabla_A - \nabla_X >0,
\label{eq_led} \end{equation}
where $\nabla_R$ and $\nabla_A$  are the radiative and adiabatic temperature gradients, respectively; and 
\begin{equation}\label{nabledoux}
\nabla_X=\sum_j \frac{\partial \ln T(\rho,p,X)}{\partial X_j}\frac{dX_j}{d\ln p}.
\end{equation} is the composition contribution to the temperature gradient.
If $\nabla_X=0$ (uniform composition), convection sets in when $\nabla_R>\nabla_A$, which is the usual \swr criterion \citep{schwarz06}. 

The temperature profile is determined by 
\begin{eqnarray}
{\frac{\partial T}{\partial m}}&=&\nabla{\frac{\partial p}{\partial m}}, \label{T_eq}
\end{eqnarray}
where $\nabla=d\ln T/d\ln p$ is the temperature gradient, which may be radiative (and conductive) or convective. 
The mixing of heat and material in convective regions is calculated according to the Mixing Length Recipe (MLR). The ratio of the mixing length to the pressure scale height is taken here to be $\alpha=\ell/H_p=0.5$, an investigation of the sensitivity of the model to a value of $\alpha$ can be found in paper~I. 
Further details of the thermal and compositional evolution calculations can be found in Appendix A of paper~I. 

\section{Computing $J_2$ and MOI}
In order to derive the second gravitational moment $J_2$ we use the standard {\it theory of figures} as described in \cite{zharktrub78}.  
For a given value of the smallness parameter $m$ and a density distribution $\rho(r)$, the level surfaces for constant internal potential can be evaluated and the surface harmonics can be computed from a series approximation in $m$ to the equation,
\begin{equation}
Ma^n\mathrm{J}_n = - \int_\tau \rho(r) r^n \mathrm{P}_n(\cos \theta) d \tau,
\label{Jn}
\end{equation}
where $a$ is the equatorial radius, and the integration is carried out over the volume $\tau$. 
$J_2$ is computed to third order; the derived values for the various models are listed in Table 1. 
\par

The moment of inertia of a spherical shell of mass $dm\,$, and radius $r\,$, relative to an axis through the center of the shell, is $(2/3)r^2dm\,$. For a spherically-symmetric
planet, the moment of inertia is therefore
\begin{equation}
\frac{2}{3}\int_M r^2 dm. 
\end{equation}
The integral is computed by summing over all the mass shells.

\bibliographystyle{apj} 
\bibliography{allona}   

\newpage

\begin{table}[ht]
\centering 
\begin{tabular}{c c c c c c c c c} 
 \hline\hline 
 &  M$_Z$ total  & Radius & T$ _{\rm eff}$ & T$_c$ & $\rho_c$ & MOI & $J_2$ & Z$_{env}$\\ [0.5ex] 
  &  [\me] &  [R$/R_p$] & [K] & [K] & [g cm$^{-3}$] & & & \\ [0.5ex] 
 \hline 

Case-J$_0$ & 40  & 1.007 & 124.6 & 1.8$\times 10^4$ & 4.3 &  0.262 & 0.01345 & 0.12\\
Case-J$_1$ & 36  & 1.001 & 124.7 & 2.5$\times 10^4$ & 10  &  0.262 & 0.01439 & 0.11\\
Case-J$_2$ & 42  & 1.004 & 124.9 & 7$\times 10^4$ & 10.6 &  0.247 & 0.01458 & 0.05\\
Case-S$_0$ & 34  & 1.003 & 94.1 & 7.3$\times 10^4$ & 5.6 & 0.228 & 0.01749 & 0.20\\
Case-S$_1$ & 35  & 0.998 & 95.2 & 7.2$\times 10^4$ & 6 & 0.228 & 0.01670 & 0.21\\
Case-S$_2$ & 36  & 0.993 & 94.9 & 3.1$\times 10^4$ & 6.7 & 0.223 & 0.01615 & 0.22\\
Case-S$_3$ & 28  & 1.004 & $^*$89.2 & 5$\times 10^4$ & 6.6 & 0.222 & 0.0166 & 0.20\\
Case-S$_4$ & 28  & 1.000 & $^*$85.7 & 6.4$\times 10^4$ & 6.2 & 0.226 & 0.0168 & 0.20\\ [1ex] 
\hline 
 \hline 
\end{tabular}
\caption{{\bf Results for Jupiter and Saturn models.} Listed are the total mass of the heavy elements (core and envelope), normalized radius, effective and central temperatures, central density, normalized moment of inertia (MOI$\equiv C/MR_{\rm eq}^2$), the second gravitational moment $J_2$, and heavy element fraction in the envelope. The values correspond to the current-state internal structure. 
The $R_p$ of Jupiter is 69,911 km and of Saturn is 58,232 km. 
For comparison, the estimated MOI values for Jupiter and Saturn are $\sim~0.26$ and $\sim 0.22$, respectively, and the measured effective temperatures are 124.4$\pm 0.3$ and 95.0$\pm 0.4$, respectively \citep[see][]{guillgaut14}. 
The derived $J_2$ values should be compared to the measured values of $J_2$(10$^{-6}$) of 14,695.62 $\pm$ 0.29 for Jupiter and 16,290.71 $\pm$ 0.27 for Saturn.
$^*$ The calculated effective temperature is very slightly affected by the energy release from the formation of the helium shell. In this model the helium shell retains heat. Instantaneous release of the settling energy could increase effective temperatures by up to 8\% for Case-S$_3$ and up to and 6\% for Case-S$_4$, see text for details.
}\label{tab1}
\end{table}

\end{document}